\def\HI{H{\small I}\ }

\def\simgt{\lower.5ex\hbox{$\; \buildrel > \over \sim \;$}}
\def\simlt{\lower.5ex\hbox{$\; \buildrel < \over \sim \;$}}

\documentclass[useAMS,usenatbib]{mn2e}

\usepackage{graphicx}

\title[The neutral hydrogen of edge-on galaxies]
  {Structure and kinematics of edge-on galaxy discs --
II. Observations of the neutral hydrogen}
\author[M. Kregel, P.~C. van der Kruit \& W.~J.~G. de~Blok]
  {M. Kregel$^1$,  
  P~.C. van~der~Kruit$^1$\thanks{E-mail: vdkruit@astro.rug.nl} 
  and W.~J.~G. de~Blok$^2$\thanks{Now at Department of Physics and 
Astronomy, Cardiff University, 5 The Parade, Cardiff CF24 3YB, UK}\\
  $^1$Kapteyn Astronomical Institute, University of Groningen,
  P.O.Box 800, 9700AV Groningen, the Netherlands\\
  $^2$Australia Telescope National Facility, PO Box 76, 
 Epping NSW 1710, Australia}
\begin{document}

\date{Accepted. Received.}

\pagerange{\pageref{firstpage}--\pageref{lastpage}} \pubyear{2004}

\label{firstpage}

\maketitle 

\begin{abstract}
We present Australia Telescope Compact Array (ATCA) and Westerbork
  Synthesis Radio Telescope (WSRT) \HI observations of 15
  edge-on spiral galaxies of intermediate to late morphological
  type. The global properties and the distribution and kinematics of
  the \HI gas are analysed and discussed. We determine the rotation
  curves using the envelope-tracing method. For 10
spiral galaxies with a stellar disc truncation we find an average
ratio of the \HI radius to the truncation radius of the stellar
disc of 1.1 $\pm$ 0.2 (1$\sigma$). 

This paper has been accepted by MNRAS and is available in pdf-format
at the following URL:\\

http://www.astro.rug.nl/$\sim $vdkruit/jea3/homepage/paperII.pdf

\end{abstract}


\end{document}